\documentclass[amsmath, amssymb,10pt,aps,prx,twocolumn,notitlepage,showpacs,superscriptaddress]{revtex4-1}
\usepackage{graphicx}
\usepackage{calc}
\usepackage{braket}
\usepackage{ulem}   
\usepackage{slashed}
\usepackage{wasysym}
\usepackage{dsfont}
\usepackage{amsthm,amsmath,amsfonts,amssymb,verbatim,color}
\usepackage{graphicx}
\usepackage{subfigure}
\usepackage{bm}
\usepackage{epsfig,slashed}
\usepackage[T1]{fontenc}
\usepackage[colorlinks=true,citecolor=blue,linkcolor=blue,urlcolor=blue]{hyperref}
\renewcommand{\b}[1]{{\mathbf{#1}}}

\begin{document}
\title{Fractional Josephson effect in nonuniformly strained graphene}

\author{Shu-Ping Lee}
\affiliation{Department of Physics, University of Alberta, Edmonton, Alberta T6G 2E1, Canada}
\author{Debaleena Nandi}
\affiliation{Department of Physics, Harvard University, Cambridge, MA 02138, USA}
\author{Frank Marsiglio}
\affiliation{Department of Physics, University of Alberta, Edmonton, Alberta T6G 2E1, Canada}
\affiliation{Theoretical Physics Institute, University of Alberta, Edmonton, Alberta T6G 2E1, Canada}
\author{Joseph Maciejko}
\affiliation{Department of Physics, University of Alberta, Edmonton, Alberta T6G 2E1, Canada}
\affiliation{Theoretical Physics Institute, University of Alberta, Edmonton, Alberta T6G 2E1, Canada}
\affiliation{Canadian Institute for Advanced Research, Toronto, Ontario M5G 1Z8, Canada}

\date{\today}

\begin{abstract}
Nonuniform strain distributions in a graphene lattice can give rise to uniform pseudomagnetic fields and associated pseudo-Landau levels without breaking time-reversal symmetry. We demonstrate that by inducing superconductivity in a nonuniformly strained graphene sheet, the lowest pseudo-Landau levels split by a pairing gap can be inverted by changing the sign of the pairing potential. As a consequence of this inversion, we predict that a Josephson $\pi$ junction deposited on top of a strained graphene sheet exhibits one-dimensional gapless modes propagating along the junction. These
gapless modes mediate single electron tunneling across the junction, giving rise to the $4\pi$-periodic fractional Josephson effect.
\end{abstract}

\maketitle
\section{INTRODUCTION}
Coupling quantum Hall states with superconductivity~\cite{BeenakkerQuantumHallJosephsonJunction,SupercurrentInQuantumHall,MonicaGrapheneSC,MonicaGrapheneSC2,GilHoSCQH} has been intensively studied recently as it provides a platform for exotic excitations such as Majorana fermions~\cite{NetanelMajoranaInQuantumHall,MajoranaInGraphene} and parafermions~\cite{JasonParafermions,RogerMongParafermion,JasonParafermions3,JasonParafermions2}. In contrast to the supercurrent in conventional Josephson junctions that is mediated by Cooper pairs, these elusive modes give rise to single electron or fractional charge tunneling, which results in a fractional Josephson effect~\cite{KitaevPWire,JasonFractionalJosephsonEffect} with flux periodicity larger than $2\pi$.

Magnetic field strengths normally required for the quantum Hall effect typically suppress proximity-induced superconductivity, which leads to challenges in attempting to couple these two phenomena. Instead of coupling superconductivity and quantum Hall states, an alternative is to couple superconductivity with quantum spin Hall states~\cite{maciejko2011}, i.e., 2D time-reversal invariant topological insulators. The topological superconductor that emerges from this coupling gives rise to Majorana modes and the fractional Josephson effect~\cite{FuKaneMajorana}. Strong spin-orbit coupling, an essential ingredient of the quantum spin Hall effect, is crucial for the occurrence of the fractional Josephson effect in this proposed realization.

\begin{figure}[t]
\includegraphics[width=8.5cm]{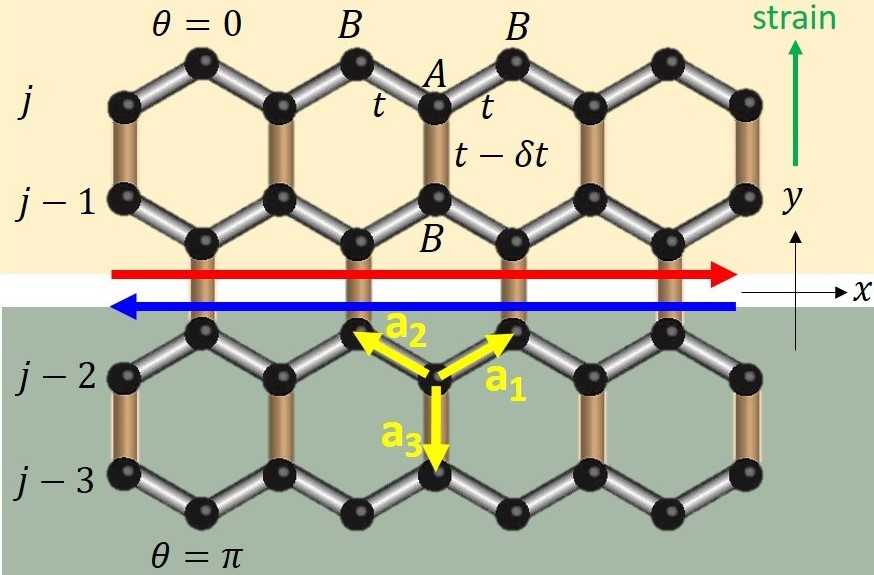}
\caption{Josephson $\pi$ junction in a graphene sheet with nonuniform uniaxial strain in the $y$ direction ($\theta$: superconducting phase, $t$: nearest neighbor electron hopping amplitude between atoms on $A$ and $B$ sublattices). We model the strain pattern by changing the hopping amplitudes on the vertical bonds (brown) by an amount $\delta t\propto y$ linearly increasing along the $y$ direction and constant along the $x$ direction. For the numerical diagonalization results (Fig.~\ref{Zigzag}) we assume the junction is parallel to the zigzag chains (gray) indexed by an integer $j$. The $\pi$ phase difference leads to an inverted pseudo-Landau level structure near the Dirac points of the graphene bandstructure. As a result, a pair of counter-propagating gapless 1D modes appears near the center of the junction (red and blue arrows).}
\label{JJStrainTightBinding}
\end{figure}

In this paper we propose an alternative way of realizing the fractional Josephson effect that requires neither spin-orbit coupling nor a magnetic field. Our proposal relies on the recently demonstrated ability to engineer large uniform pseudomagnetic fields in graphene by applying nonuniform distributions of strain~\cite{GeimStrainedGraphene,GuineaStrainedGrapheneExp,StrainTightBindingAshvin,Guinea,
MasirPseudofieldStrainedGraphene,Tony,StrainOddSCGraphene}. Just like a physical magnetic field in a 2D electron gas gives rise to the Hall effect and, upon quantization, Landau levels, in strained graphene the pseudomagnetic field leads to the valley Hall effect and pseudo-Landau levels~\cite{Vaezi}. Experimentally, strain-induced pseudomagnetic fields in excess of 300 Tesla have been reported in graphene nanobubbles~\cite{GuineaStrainedGrapheneExp,Gomes}. Strain-induced pseudo-Landau levels have also been observed in graphene grown by chemical vapor deposition~\cite{YehNaiChang}.

Coupling superconductivity to strained graphene modifies the pseudo-Landau levels~\cite{BrunoSCInStrainedGraphene} in a manner similar to how the Landau levels of a massless Dirac Hamiltonian are affected by the addition of a Dirac mass. In the Dirac-Landau problem a mass domain wall (i.e., an interface at which the Dirac mass changes sign) gives rise to a gapless chiral 1D mode that propagates along the domain wall~\cite{callan1985,qi2008} and disperses in the otherwise empty intra-Landau-level bulk gap. Likewise here, we find that an interface at which the pairing term changes sign, i.e., a Josephson $\pi$ junction, is accompanied by 1D gapless modes dispersing within the bulk pairing gap between the two lowest (zeroth) pseudo-Landau levels. These gapless modes in turn lead to a $4\pi$ energy-phase relation and the fractional Josephson effect.

\section{MODEL OF STRAINED GRAPHENE}

The device we consider is a Josephson junction between two $s$-wave superconducting electrodes deposited on top of a graphene sheet with nonuniform (linearly increasing) uniaxial strain (Fig.~\ref{JJStrainTightBinding}), for instance using the methods described in Ref.~\onlinecite{heersche2007}. There are three vectors, $\b{a}_1=a(\sqrt{3}/2,1/2)$, $\b{a}_2=a(-\sqrt{3}/2,1/2)$ and $\b{a}_3=a(0,-1)$ that connect any atom on an $A$ sublattice of the graphene lattice to the nearest neighboring atom on a $B$ sublattice, where $a$ is the closest distance between carbon atoms. In strained graphene, we assume the hopping amplitudes along $\b{a}_1$ and $\b{a}_2$ remain unmodified with value $t$, but the hopping amplitude along the $\b{a}_3$ direction is reduced as $t-\delta t(y)$~\cite{StrainTightBindingAshvin,StrainTightBinding} where $\delta t(y)\propto y$ represents the effect of linearly increasing strain in the $y$ direction. The Hamiltonian with nearest neighbor hopping is composed of two terms, $H=H_0+H_s$. The first is the usual nearest neighbor hopping term on the honeycomb lattice,
\begin{eqnarray}
H_0&=&-t \sum_{{\bf{r}}\sigma} \sum^{3}_{j=1}c^{\dagger}_{B,\b{r}+\b{a}_j,\sigma}c_{A,{\bf{r}}\sigma}+\mathrm{H.c.},
\label{TightBindingHamiltonianRealSpace}
\end{eqnarray}
where $\b{r}$ denotes sites of the underlying triangular Bravais lattice and $\sigma=\uparrow,\downarrow$ is spin. The second represents a modification due to strain, and is given by
\begin{eqnarray}
H_s&=&\sum_{{\bf{r}}}\delta t(y) c^{\dagger}_{B,{\b{r}+\b{a}_3}}c_{A,{\bf{r}}}+\mathrm{H.c.}
\label{StrainHamiltonianRealSpace}
\end{eqnarray}

Eq.~(\ref{TightBindingHamiltonianRealSpace}) can be expressed in momentum space as
\begin{align}
H_0=-t\sum_{\b{k}\sigma}\sum_{m=1}^3 e^{i{\bf{k}}\cdot {\b{a}_m}}c^{\dagger}_{B\sigma}({\bf{k}})c_{A\sigma}({\bf{k}})+\mathrm{H.c.}
\end{align}
Expanding the momentum ${\bf{k}}$ near the Dirac points $K_{\pm}\equiv\pm4\pi/(3\sqrt{3}a)(1,0)$ as ${\bf{k}}=K_{\pm}+{\bf{p}}$ where $|\b{p}|$ is small compared to the dimensions of the Brillouin zone allows us to rewrite $H_0$ as a Dirac Hamiltonian~\cite{GeimRMP,StrainTightBindingAshvin,ChangYuHou} in the basis of $\Psi_\sigma({\bf{p}})=(c^{K_{+}}_{A\sigma}({\bf{p}}),c^{K_{+}}_{B\sigma}({\bf{p}}),c^{K_{-}}_{A\sigma}({\bf{p}}),
c^{K_{-}}_{B\sigma}({\bf{p}}))^T$ where $c_{A,B,\sigma}^{K_\pm}(\b{p})\equiv c_{A,B,\sigma}(K_\pm+\b{p})$. More specifically, the effective low-energy Hamiltonian near the Dirac points is expressed as $H_0\approx\sum_\sigma\int [d^2p/(2\pi)^2]\Psi_\sigma^{\dagger}\mathcal{H}_0\Psi_\sigma$, where
\begin{equation}
\mathcal{H}_0=v_F(\tau_z\sigma_x p_x+\sigma_y p_y),
\label{Hamiltonian}
\end{equation}
describes linearly dispersing Dirac fermions with velocity $v_F\equiv(3/2)at$. We use $\sigma_i$ and $\tau_i$, $i=x,y,z$ to denote the Pauli matrices in sublattice ($A$ and $B$) and valley ($K_+$ and $K_-$) space, respectively.

We now turn to the term $H_s$ in Eq.~(\ref{StrainHamiltonianRealSpace}) that is due to strain. Assuming $\delta t(y)$ varies slowly on the scale of the lattice constant $a$, the two valleys remain approximately decoupled and one can write $H_s\approx H^{K_{+}}_s+H^{K_{-}}_s$ with~\cite{StrainTightBindingAshvin,StrainTightBinding,MasirPseudofieldStrainedGraphene}
\begin{align}\label{StrainHamiltonContinuous}
H^{K_{\pm}}_s =\sum_\sigma\int d^2r\,\delta t(y)e^{iK_{\pm} \cdot \b{a}_3} c^{K_\pm}_{B\sigma}(\b{r})^\dag c^{K_\pm}_{A\sigma}(\b{r})+\mathrm{H.c.},
\end{align}
where $\b{r}=(x,y)$ now denotes the continuum position and $c_{A,B,\sigma}^{K_\pm}(\b{r})$ is the continuum Fourier transform of $c_{A,B,\sigma}^{K_\pm}(\b{p})$. In our choice of basis $K_\pm\cdot\b{a}_3=0$, thus the strain Hamiltonian becomes $H_s\approx\sum_\sigma\int d^2r\,\Psi_\sigma^\dag\mathcal{H}_s\Psi_\sigma$ with
\begin{equation}
\mathcal{H}_s=\delta t(y)\sigma_x.
\label{StrainHamiltonContinuous2}
\end{equation}
The full Hamiltonian for nonuniformly strained graphene in the continuum limit is given as
\begin{equation}
\mathcal{H}=\mathcal{H}_0+\mathcal{H}_s=v_F\left[\tau_z\sigma_x \left(-i\partial_x+\delta t(y) \tau_z/v_F\right)-i\sigma_y \partial_y\right].
\label{StrainGrapheneContinuous3}
\end{equation}
Comparing Eq.~(\ref{Hamiltonian}) and Eq.~(\ref{StrainGrapheneContinuous3}), the variation in hopping strength along the $\b{a}_3$ direction is equivalent to the substitution $-i
\partial_x\rightarrow -i\partial_x+\delta t(y) \tau_z/v_F$. In other words, the strain in graphene generates a pseudo-vector potential~\cite{StrainTightBindingAshvin,StrainTightBinding} along the $x$ direction given by ${\b{A}_s}(\b{r})=\tau_z\delta t(y)\hat{\b{x}}/(ev_F)$. The Pauli matrix $\tau_z$ in ${\b{A}_s}$ signifies that electrons in different valleys experience a pseudomagnetic field of the same magnitude but opposite sign. This comes from the fact that the
pseudomagnetic field preserves time-reversal symmetry. One can indeed check that  Eq.~(\ref{StrainGrapheneContinuous3}) is invariant under a time-reversal transformation $\mathcal{T}$. Momentum changes sign under time reversal, which implies $\mathcal{T}\tau_z\mathcal{T}^{-1}=-\tau_z$ because we exchange the valley $K_{+}$ to $-K_{+}=K_{-}$ (modulo a reciprocal lattice vector) under time reversal. The sublattices $A$ and $B$ remain the same under time reversal, giving $\mathcal{T}\sigma_x\mathcal{T}^{-1}=\sigma_x$ and $\mathcal{T}\sigma_y\mathcal{T}^{-1}=-\sigma_y$ because
$\mathcal{T}i\mathcal{T}^{-1}=-i$. Combining these results gives $\mathcal{T}\mathcal{H}\mathcal{T}^{-1}=\mathcal{H}$.

Because a linearly increasing vector potential corresponds to a uniform magnetic field, we expect pseudo-Landau levels to appear in Eq.~(\ref{StrainGrapheneContinuous3}). Writing the strain-induced variation in hopping strength as $\delta t(y)=ev_FBy$ where $B$ is the pseudomagnetic field~\cite{Pesin}, Eq.~(\ref{StrainGrapheneContinuous3}) becomes
\begin{equation}
\mathcal{H}=v_F[\tau_z\sigma_x (p_x+\tau_z eBy)-i\sigma_y\partial_y].
\label{HamiltonianInPseudoMagneticField}
\end{equation}
Translation invariance in the $x$ direction allows us to replace the momentum operator in the $x$ direction by its eigenvalue $p_x$, while translation invariance is broken by the strain pattern in the $y$ direction. Equation~(\ref{HamiltonianInPseudoMagneticField}) is almost the same as the Dirac Hamiltonian in a uniform magnetic field in the Landau gauge, except for the presence of $\tau_z$ in the vector potential. To demonstrate the presence of pseudo-Landau levels, we define the dimensionless coordinate $\xi \equiv l_Bp_x+y/l_B$ for the $K_{+}$ valley and $\overline{\xi} \equiv l_Bp_x-y/l_B$ for the $K_{-}$ valley, where $l_B \equiv 1/\sqrt{eB}$ is a pseudomagnetic length. This allows us to define the raising operator as $a^{\dagger} \equiv \frac{1}{\sqrt{2}}(\xi-\partial_{\xi})$ for the $K_{+}$ valley and $\overline{a}^{\dagger} \equiv \frac{1}{\sqrt{2}}(\overline{\xi}-\partial_{\overline{\xi}})$ for the $K_{-}$ valley. The corresponding lowering operators are $a=\frac{1}{\sqrt{2}}(\xi+\partial_{\xi})$ and $\overline{a}=\frac{1}{\sqrt{2}}(\overline{\xi}+\partial_{\overline{\xi}})$. These operators satisfy the commutation relations $[a,a^{\dagger}]=1$, $[\overline{a},\overline{a}^{\dagger}]=1$, and allow us to write the first quantized Hamiltonian (\ref{HamiltonianInPseudoMagneticField}) in a simple form,
\begin{equation}
\mathcal{H}=\frac{\sqrt{2}v_F}{l_B}
\begin{pmatrix}
    0 & a^{\dagger} & 0 & 0\\
    a & 0 & 0 & 0\\
    0 & 0 & 0 & -\overline{a}^{\dagger}\\
    0 & 0 & -\overline{a} & 0
\end{pmatrix}.
\label{HarmonicOscillatorsInPseudoMagneticField}
\end{equation}
By solving the eigenvalue problem $\mathcal{H}\psi_n=E_n\psi_n$, we obtain the pseudo-Landau levels as $E_n=\pm \sqrt{2}v_F\sqrt{n}/l_B$, where $n=0,1,2,\ldots$ The corresponding eigenstates $\psi_n=(\psi^{K_{+}}_n,\psi^{K_{-}}_n)^T$ are given in terms of the simple harmonic oscillator eigenstates $|n\rangle$ as $\psi^{K_{+}}_n=(|n\rangle,\pm|n-1\rangle)^T$ and $\psi^{K_{-}}_n=(|\overline{n}\rangle,\mp|\overline{n-1}\rangle)^T$ for the $K_{+}$ and $K_{-}$ valleys, respectively. Here the simple harmonic oscillator eigenstates $|n\rangle$ are given as $|n\rangle\propto\exp(-\xi^2/2)H_n(\xi)$ for the $K_{+}$ valley and $|\overline{n}\rangle\propto\exp(-\overline{\xi}^2/2)H_n(\overline{\xi})$ for the $K_{-}$ valley, where $H_n$ is the $n$th Hermite polynomial. The wavefunctions are plane waves in the $x$ direction, with an infinite degeneracy parametrized by the momentum eigenvalue $p_x$.

The wavefunction of the zeroth pseudo-Landau level is $\psi^{K_{+}}_0=(|0\rangle,0)^T$ for the $K_{+}$ valley and $\psi^{K_{-}}_0=(|\overline{0}\rangle,0)^T$ for the $K_{-}$ valley, thus the wavefunctions for both valleys have support only on the A sublattice. This is in contrast to the case of a real magnetic field where the zeroth Landau level wavefunctions for the $K_{+}$ and $K_{-}$ valleys have support on the A and B sublattice, respectively. Support on the same sublattice in a pseudomagnetic field is one of the key reasons why an on-site pairing potential opens a superconducting gap in the zeroth pseudo-Landau level, as will be seen shortly.

\section{PROXIMITY-INDUCED SUPERCONDUCTIVITY}

To couple the pseudo-Landau levels with superconductivity, we add a proximity-induced pairing term $\Delta$ to the Hamiltonian,
\begin{eqnarray}
H_{\Delta}=&&\sum_{\alpha=A,B}\int\frac{dp_x}{2\pi} \left(\Delta c^{K_{+}\dagger}_{\alpha\uparrow}(p_x) c^{K_{-}\dagger}_{\alpha\downarrow}(-p_x)+\mathrm{H.c.}\right)\nonumber\\
&&+(\uparrow\, \leftrightarrow\, \downarrow),
\end{eqnarray}
using $\uparrow, \downarrow$ to denote the physical spin. The full Hamiltonian with the pairing term included is $SU(2)$ spin rotationally invariant and all eigenstates have an exact two-fold degeneracy; in the following we factor out this degeneracy and focus on solving the Hamiltonian in one of the two nondegenerate subspaces. This Hamiltonian can be expressed in the Nambu/Bogoliubov-de Gennes (BdG) basis $\Psi^{\text{BdG}}(p_x)=[c^{K_{+}}_{A\uparrow}(p_x),c^{K_{+}}_{B\uparrow}(p_x),c^{K_{-}\dagger}_{A\downarrow}(-p_x),c^{K_{-}\dagger}_{B\downarrow}(-p_x)]^T$ as
\begin{eqnarray}
\mathcal{H}_{\text{BdG}}=\frac{v_F}{\sqrt{2}l_B}
\begin{pmatrix}
    0 & a^{\dagger} & \tilde{\Delta} & 0\\
    a & 0 & 0 & \tilde{\Delta}\\
    \tilde{\Delta}^{*} & 0 & 0 & -a^{\dagger}\\
    0 & \tilde{\Delta}^{*} & -a & 0
\end{pmatrix},
\label{BdGInPseudoMagneticField}
\end{eqnarray}
where $\tilde{\Delta}=\frac{\sqrt{2}l_B}{v_F}\Delta$ is a dimensionless measure of the pairing gap. Diagonalizing Eq.~(\ref{BdGInPseudoMagneticField}), we obtain the energy spectrum as $E_n=\pm\frac{v_F}{\sqrt{2}l_B}\tilde{E}_n$ where $\tilde{E}_n\equiv\sqrt{|\tilde{\Delta}|^2+n}$ is dimensionless. The corresponding (unnormalized) eigenstate is given as $\psi^{\text{BdG}}_n=(\tilde{\Delta} |n\rangle,\tilde{\Delta} |n-1\rangle,(\tilde{E}_n-\sqrt{n})|n\rangle,(\tilde{E}_n-\sqrt{n})|n-1\rangle)^T$. The energy spectrum $E_n$ is the same as the Landau level spectrum of a massive Dirac Hamiltonian with mass $\tilde{\Delta}$. In a 2D topological insulator, a change of sign of the mass in the Dirac Hamiltonian is accompanied by band inversion. We thus expect the zeroth pseudo-Landau level to become inverted as the sign of the pairing term in Eq.~(\ref{BdGInPseudoMagneticField}) is reversed.
\begin{figure}
\includegraphics[width=8.5cm]{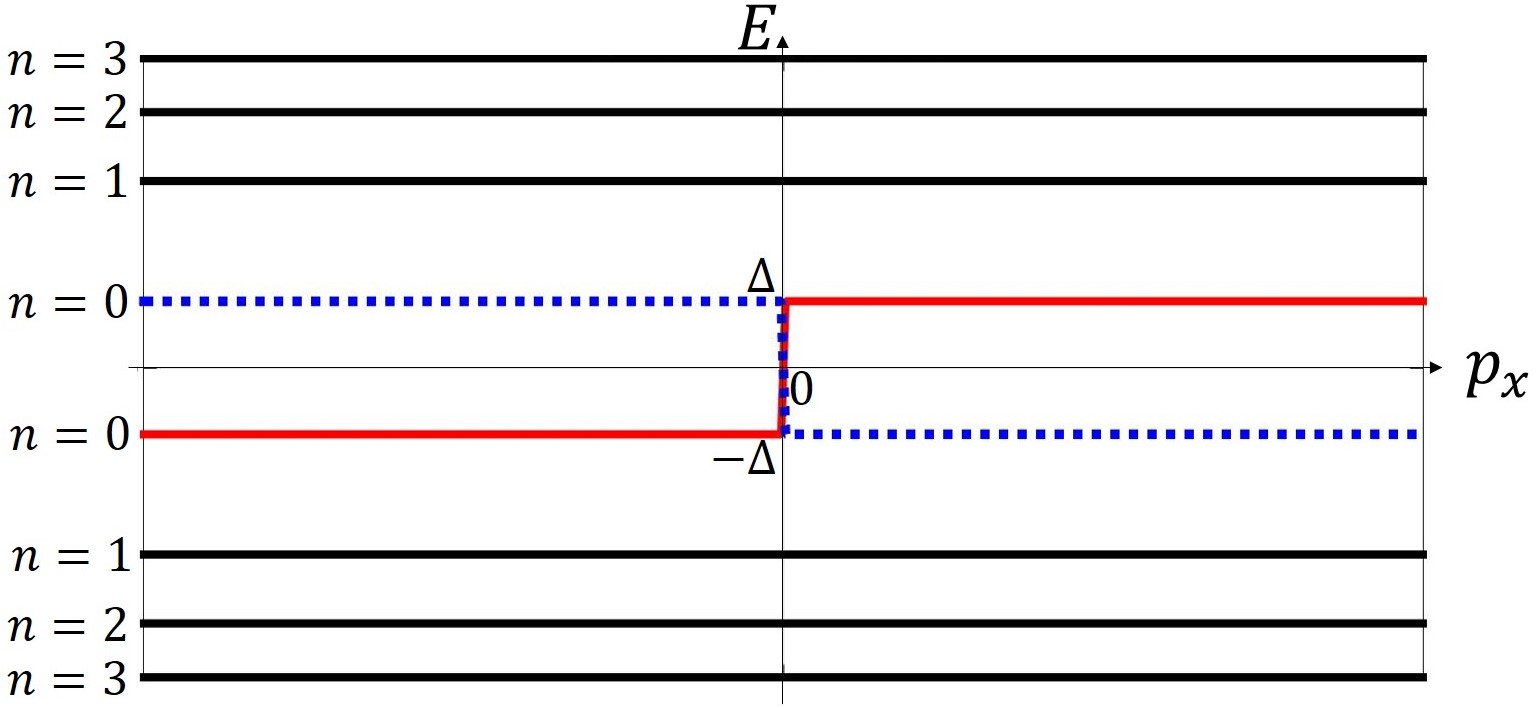}
\caption{Bogoliubov-de Gennes spectrum for the continuum model of strained graphene with a Josephson $\pi$ junction built on top, with $n$ the pseudo-Landau level index and $p_x$ the momentum along the junction. Assuming the junction is at $y=0$, the superconducting phase varies from $\theta=0$ when $y>0$ to $\theta=\pi$ when $y<0$. The guiding center of the pseudo-Landau level is located at $y_0=-l^2_Bp_x$, which moves from the $\theta=0$ region to the $\theta=\pi$ region as $p_x$ is varied from negative to positive. The $n=0$ pseudo-Landau level becomes inverted if either the pairing potential or $p_x$ change sign.}
\label{BandInversion}
\end{figure}

We verify this intuition by explicit calculation. The $n=0$ pseudo-Landau level splits into dispersionless BdG bands $E_0=\pm|\Delta|$ in the presence of the pairing gap $|\Delta|$. Writing the pairing potential as $\Delta=|\Delta| e^{i\theta}$, the corresponding wavefunction $\psi(\theta,E_0)$ of the zeroth pseudo-Landau level with energy $E_0$ and phase $\theta$ is
\begin{eqnarray}
\psi_0^{\text{BdG}}(\theta,E_0=+|\Delta|)&=&(|0\rangle,0,+e^{-i\theta}|0\rangle,0)^T,\nonumber\\
\psi_0^{\text{BdG}}(\theta,E_0=-|\Delta|)&=&(|0\rangle,0,-e^{-i\theta}|0\rangle,0)^T.
\label{LLLWavefunction}
\end{eqnarray}
The zeroth pseudo-Landau level wavefunction with energy $E_0=|\Delta|$ and phase $\theta=0$ is the same as the wavefunction with energy $E_0=-|\Delta|$ and phase $\theta=\pi$. In other words,
\begin{align}
\psi_0^{\text{BdG}}(\theta=0,E_0=+|\Delta|)&=\psi_0^{\text{BdG}}(\theta=\pi,E_0=-|\Delta|),\nonumber\\
\psi_0^{\text{BdG}}(\theta=0,E_0=-|\Delta|)&=\psi_0^{\text{BdG}}(\theta=\pi,E_0=+|\Delta|).
\label{BandInversionEq}
\end{align}
Equation (\ref{BandInversionEq}) is the main result of this work, and means that the zeroth pseudo-Landau level undergoes band inversion as the sign of the pairing potential is reversed. In experiment, reversing the sign of the pairing potential can be achieved by building a Josephson junction with a phase difference of $\pi$ across the junction. For this reason, we consider a Josephson junction that is built on top of strained graphene as shown in Fig.~\ref{JJStrainTightBinding}.

We now analyze the energy-momentum relation of the zeroth pseudo-Landau level in a Josephson $\pi$ junction. The zeroth pseudo-Landau level wavefunction is $|0\rangle\propto\exp(-\xi^2/2)$, which peaks at $\xi=0$. By setting $\xi=0$, we locate the guiding center of the wavefunction at $y_0=-l^2_Bp_x$. Assuming the pairing potential changes sign at $y=0$, the pseudo-Landau levels at energy $\pm|\Delta|$ cross as the momentum $p_x$ goes from negative to positive (Fig.~\ref{BandInversion}).

\begin{figure}
\includegraphics[width=\columnwidth]{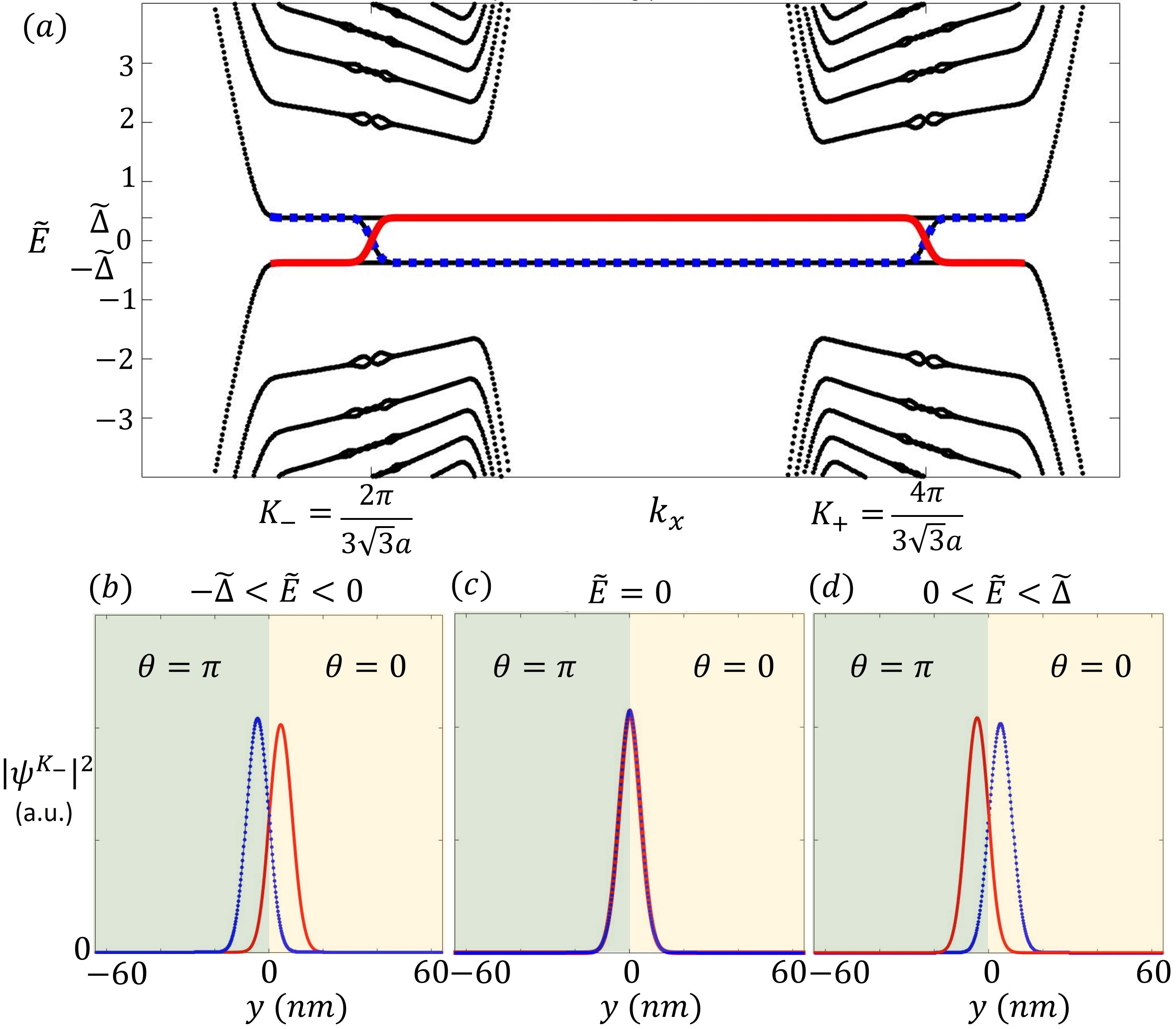}
\caption{(a) Bogoliubov-de Gennes spectrum for the lattice model of a strained graphene strip (600 sites wide) with zigzag edges and a Josephson $\pi$ junction built on top, as a function of the momentum $k_x$ along the junction centered at $y=0$. The chosen strain field is equivalent to a $20$~T pseudomagnetic field, and the pairing gap is $1\%$ of the nearest neighbor hopping strength. $\tilde{E}_n$ is the energy in units of $v_F l_B^{-1}/\sqrt{2}$. The zeroth pseudo-Landau level is inverted at both the $K_{+}$ and $K_{-}$ valleys, with gapless chiral 1D modes inside the pairing gap $\tilde{\Delta}$ appearing along the $\pi$ Josephson junction in both valleys. We plot the probability density for the gapless modes at the $K_-$ valley along the $y$ direction when the energy is (b) below the Dirac point, (c) at the Dirac point and (d) above the Dirac point.}
\label{Zigzag}
\end{figure}

Figure~\ref{BandInversion} implies that when the zeroth pseudo-Landau level is inverted across the $\pi$ junction, gapless modes appear within the pairing gap. We confirm this result achieved with a continuum model with that obtained from a lattice model. The energy-momentum relation of strained graphene with a $\pi$ junction in a lattice model is shown in Fig.~\ref{Zigzag}(a). We consider a graphene lattice with translation invariance in the $x$ direction but with a finite width with zigzag edges in the
$y$ direction as in Fig.~\ref{JJStrainTightBinding}. We Fourier transform Eq.~(\ref{TightBindingHamiltonianRealSpace}) and Eq.~(\ref{StrainHamiltonianRealSpace}) along the $x$ direction to wavevector space $k_x$, which gives the tight-binding Hamiltonian of a strained graphene strip,
\begin{align}
H=&\sum_\sigma\int\frac{dk_x}{2\pi}\Biggl[-2t\cos\frac{\sqrt{3}k_xa}{2}
\sum_{j=1}^{N}c^{\dagger}_{B,j\sigma}(k_x)c_{A,j\sigma}(k_x)\nonumber\\
&-\sum_{j=2}^{N}\bigl(t-\delta t(y)\bigr)c^{\dagger}_{B,j-1,\sigma}(k_x)c_{A,j\sigma}(k_x)+\textrm{H.c.}\Biggr],
\label{StrainedTightBindingHamiltoniankx}
\end{align}
where $j=1,\ldots,N$ is the lattice index along the $y$ direction and $y\equiv(3/2)a[j-(N+1)/2]$ in $\delta t(y)$. The proximity-induced pairing term becomes
\begin{align}
H_{\Delta}=&\sum_{\alpha=A,B}\sum_{j=1}^N\int\frac{dk_x}{2\pi}\left(\Delta c^{\dagger}_{\alpha j\uparrow}(k_x) c^{\dagger}_{\alpha j\downarrow}(-k_x)+\mathrm{H.c.}\right)\nonumber\\
&+(\uparrow\,\leftrightarrow\,\downarrow).
\end{align}

To model a Josephson $\pi$ junction, the pairing potential is expressed as $\Delta=\Delta_0 e^{i\theta}$ where $\theta=0$ for $y<0$ and $\theta=\pi$ when $y>0$. Diagonalizing the total Hamiltonian $H+H_{\Delta}$ numerically, we find a crossing of the zeroth pseudo-Landau levels at the both the $K_+$ and $K_-$ Dirac points [Fig.~\ref{Zigzag}(a)], consistent with our analysis of the continuum model (Eq.~(\ref{BandInversionEq}) and Fig.~\ref{BandInversion}). This crossing results in gapless 1D propagating modes localized near the $\pi$ junction [Fig.~\ref{Zigzag}(b)-(d)] and appearing within the pairing gap. For energies near the Dirac point $|\tilde{E}|\ll|\tilde{\Delta}|$, the gapless modes disperse linearly with momentum $\tilde{E}\propto\pm p_x$. Since the guiding center of the zeroth pseudo-Landau wavefunctions is $-l_B^2p_x$, in this low-energy regime the spatial separation between the two counter-propagating modes is proportional to $\tilde{E}$.

\begin{figure}
\includegraphics[width=8.5cm]{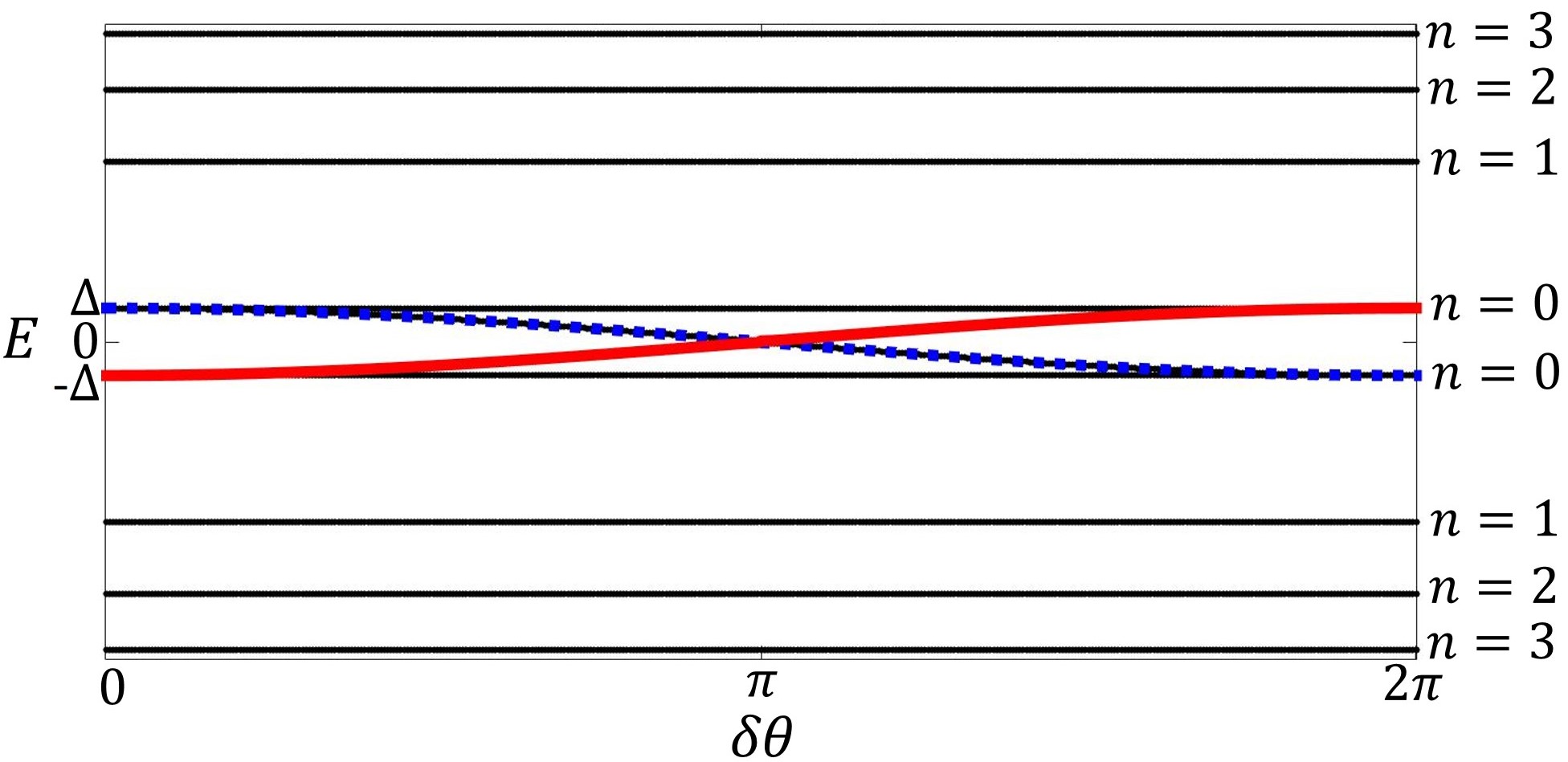}
\caption{Bogoliubov-de Gennes spectrum at the Dirac point of a Josephson junction on a strained graphene strip as a function of the phase difference $\delta\theta$ across the junction, computed from the lattice model. The zeroth pseudo-Landau level exhibits a $4\pi$-periodic energy-phase relation, leading to a $4\pi$-periodic Josephson supercurrent. The other pseudo-Landau levels do not disperse with $\delta\theta$ and thus do not contribute to the supercurrent.}
\label{EnergyPhase}
\end{figure}

\section{DISCUSSION}

The 1D gapless modes mediate single electron tunneling, which leads to a fractional Josephson effect with a $4\pi$ energy-phase relation (Fig.~\ref{EnergyPhase}). The pseudo-Landau level inversion in the $\pi$ junction implies a zero-energy mode at the $K_{+}$ and $K_{-}$ points. Denoting by $\delta \theta$ the superconducting phase difference across the junction, the zero-energy mode $E(\delta \theta=\pi)=0$ at the $K_{+}$ or $K_{-}$ Dirac point implies $4\pi$ periodicity in the energy-phase relation of the junction. To demonstrate this effect, we fix the momentum $k_x=\pm4\pi/(3\sqrt{3}a)$ at one of the Dirac points $K_{\pm}$ and diagonalize the total Hamiltonian $H+H_{\Delta}$ while varying the phase difference $\delta\theta$. This gives the spectrum shown in Fig.~\ref{EnergyPhase}. Most of the pseudo-Landau levels do not disperse with $\delta\theta$, except the zeroth level which has a $4\pi$-periodic energy-phase relation. The supercurrent $I_s\propto\partial E/\partial (\delta\theta)$ is proportional to the derivative of the total energy $E$ with respect to the phase difference $\delta\theta$, which means only the zeroth pseudo-Landau level contributes, giving rise to a $4\pi$-periodic current-phase relation. In the absence of a $\pi$ junction, the non-dispersing pseudo-Landau levels correspond to a flat band superconductor~\cite{FlatBandSuperconductivity}.

The $4\pi$-periodic Josephson effect is traditionally associated with unpaired Majorana fermions in 1D topological superconductors~\cite{KitaevPWire}. While in the topological superconductor two Majorana modes at opposite ends of a wire form a single zero-energy electronic state which mediates single-electron tunneling, in our case this is achieved by two zero-energy electronic states in each valley which are degenerate by $SU(2)$ spin rotation symmetry. The zero-energy modes at the $K_+$ and $K_-$ Dirac points [see Fig.~\ref{Zigzag}(a)] can be expressed in terms of eight Majorana operators. Unlike in topological superconductors, however, such Majorana modes are not topologically protected.

In conclusion, we have shown that the zeroth pseudo-Landau level induced by straining graphene in zero magnetic field can be inverted by a Josephson $\pi$ junction, which reverses the sign of the pairing potential. As a consequence of pseudo-Landau level inversion, a pair of gapless counter-propagating 1D modes appears near the center of the junction. These gapless modes in turn lead to a $4\pi$ energy-phase relation, giving rise to the possibility of observing the fractional Josephson effect in strained graphene.

\acknowledgments{
We thank Chien-Hung Lin for illuminating discussions. This research was supported by NSERC grants \#RGPIN-203396 (F.M.) and \#RGPIN-2014-4608 (J.M.), the Canada Research Chair (Program CRC), the Canadian Institute for Advanced Research (CIFAR), and the University of Alberta.}

%

\end{document}